\documentclass[aps,pra,twocolumn,longbibliography]{revtex4-2}

\usepackage{amsmath,amssymb}
\usepackage{mathrsfs}
\usepackage{graphicx}
\usepackage{physics}
\usepackage{xcolor}
\usepackage{dsfont}
\usepackage[hidelinks]{hyperref}
\usepackage[normalem]{ulem}
\usepackage[utf8]{inputenc}
\usepackage[T1]{fontenc}
\usepackage[protrusion=true, expansion=true]{microtype}

\hypersetup{colorlinks=true, linkcolor=black, citecolor=black, urlcolor=blue}

\begin{document}

%\title{Non-equilibrium dynamics and correlations of exotic states in Transmon Pairs}
%\title{Dynamics of Dark States in Weakly Anharmonic Transmon Arrays}
\title{Perturbative Analysis of Dark State Dynamics in Weakly Anharmonic Photon-Emitter Pairs}
%\title{Dynamics of Dark States in WEakly Anharmonic Transmon Arrays}

\author{Christopher Campbell}
\email[Corresponding author: ]{christopher.campbell@oulu.fi}
\author{Matti Silveri}
\affiliation{Nano and Molecular Systems Research Unit, University of Oulu, Oulu Finland} 
\date{\today}

\begin{abstract}

Dark states are excited quantum states that decouple from their environment in such a way that they do not emit or absorb external photons. These states are found in a variety of different open quantum systems and can be derived from the collective interactions of individual quantum emitters interacting with one another. One of the simplest model where these states exist is in a pair of dissipatively coupled harmonic oscillators described under the Bose-Hubbard model. When on-site interactions are included, these states can no longer be classified as genuine dark states since dissipation is induced in them. In this paper we study the origin of this dissipation in dark states by using weak anharmonicity as a perturbing factor. In our analysis, we find the first and second order corrections to the wavefunction and apply these corrections to the master equation in order to track the dynamics. 

\end{abstract}

\maketitle

\section{Introduction}

Photon-emitters that collectively interact with one another through a medium such as a waveguide or cavity have the potential to decouple from an environment to produce a decoherence free space. In this particular form interaction, a quantum state manifests as a superposition of interacting Fock states where photons are neither emitted nor absorbed by the system, known as a dark state~\cite{Finkelstein2019,Tokman2023,Celso2025}. This particular state can serve as a resource in quantum information retention and computing operations~\cite{Tokman2025,Petrosyan2017}, since by decoupling from the environment information is protected while also exhibiting long-range entanglement~\cite{Shen2025,zou2022,Lidar1998}. When prepared these states can be robust~\cite{Rubies2022}, however, changes in interactions between the system and an environment can compromise their stability over time. Depite these set backs, methods to retain dark state information and improve their lifetimes are a popular topic~\cite{Zhou2017,Starek_2020}.

\begin{figure}[t]
    \centering
    \includegraphics[width =\linewidth]{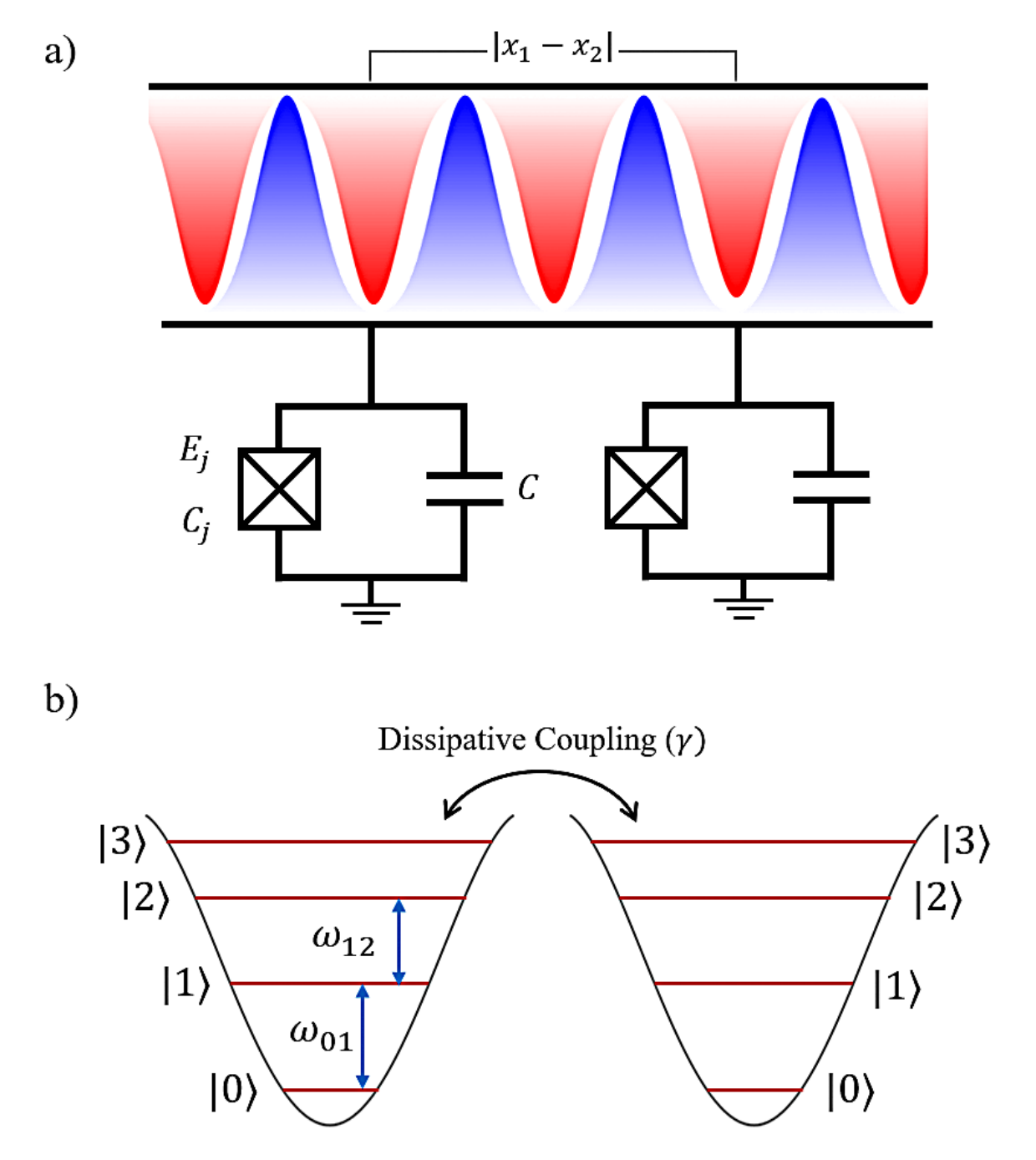}
    \caption{Schematic of the considered system of a two transmons coupled to a waveguide. (a) Two transmons separated by a distance of $d=|x_1-x_2|$. At each site a microcircuit is comprised of an LC circuit and a Josephson junction which interacts to an open transmission line or waveguide. The waveguide is open ensuring photons emitted into the environment are lost from the system. (b) Analogous system described by the Bose-Hubbard model as two anharmonic oscillators coupled via dissipative coupling. Superpositions of the Fock states between the oscillators can be observed from the collective constructive and destructive interference of the transmons, leading to the creation of dark and bright states.  }
    \label{fig:Schematic}
\end{figure}

Dark states, and the quantum optical properties of dark states, are demonstrated in many different quantum systems~\cite{Zhao2026}. In ultracold atomic gases and atomic vapors dark states serve as an excellent model to study coherent population trapping~\cite{Arimondo1976,Hemmerich1995} and stimulated Raman adiabatic passage protocols in $\Lambda$ transition configurations due to the inhibition of photon absorption and emission~\cite{Sevin2011}. Early experimental works demonstrating similar suppressions of photon absorption and emission  demonstrate electromagnetically induced transparency in dark state polaritons, where light pulses become trapped in a system through light matter interactions~\cite{HU2017,EITReview,Weatherill_2008}.

Macroscopic structures such as transmons, microcircuits that exhibit quantum mechanical properties, also have the ability to hold dark states~\cite{Holzinger2022,Zanner2022}. When  coupled to a waveguide or transmission line, they collective interact with one another, as seen in Fig.~\ref{fig:Schematic}(a), dark states appear due to the dissipative transfer of excitations when coupled to a waveguide. Since this dark state is inherently an entangled state, the transmon device has garnered interest as a platform for studying quantum information~\cite{Kwon2021,Baumer2024}. Transmons however are versatile in their applications due to the tunability of coupling parameters and anharmonicity. This also makes them a viable platform for studying many body dynamics at microscopic scales. 

In the theoretical description of their spectral properties, transmon arrays can be modeled using the effective Bose-Hubbard model as anharmonic emitters as seen in Fig.~\ref{fig:Schematic}(b)~\cite{BHReview2008,Wiegand2020,Orell2022,Olli2022,Vaaranta2022,Zanner2022}. The Hamiltonian itself is Hermitian however inclusion of dissipative components introduces couplings of the sub-systems, leading to a non-Hermitian description of the dynamics~\cite{Sergi2012,Yunzhao2023,Dong2025}. The inclusion of non-linear terms such as on-site interactions further complicates the dynamics and can lead to numerical instabilities~\cite{Feng2025}. This is largely due to the introduction of biorthogonal wavefunctions, exceptional points and wavefunction coalescence which emerge even for small sub-systems such as transmon pairs~\cite{BirothoQM2005,Brody_2014}. Mathematical methods such as Gram-Schmidt orthogonalization among others can circumnavigate effects such as biorthogonality~\cite{Toth2015,Beresford1985,Edvardsson2024}, however meaningful information related to the population of states in higher level excitation manifolds can be lost due to renormalization corrections of these wavefunctions~\cite{Zhong2025}. Because of this, finding alternative approximation methods to test against the these orthogonalization methods can prove useful, especially if an alternative method gives insight into the composition of the wavefunctions which strengthens decoherence free spaces for higher excitations past the qubit model. In this paper we will exam the effects of weak anharmonicity as a perturbation to a collectively interacting emitter pair using the Bose-Hubbard model and show how applications of first and second order corrections of dark state wavefunctions can be applied in time-independent and dependent cases with good accuracy and stability.

Research into the fundamental nature of these states as open quantum systems, collective interactions and their responses to lossy environments has provided insights into the modeling of non-Hermitian Hamiltonians. In recent studies, the application of non-degenerate perturbation theory to open quantum systems has revealed how couplings and correlations emerge in steady state light emitters, dark states, and how the application of these methods provide a simplified description of non-Hermitian physics in collectively interacting spin systems~\cite{Dolf2023}. These findings provide a starting point for the work we present in this article, where we extend perturbation methods to non-Hermitian Hamiltonians~\cite{Sternheim1972} and wavefunctions and apply the corrections a weak anharmonicity as a perturbation to the master equation. Subsequently these perturbations can track the emerging dynamics of dark states that are directly influenced by anharmonicity. By extending the methods to include higher order states in the context of transmon arrays and anharmonic oscillators, we hope that the effects of anharmonicity on entangled states can be studied in order to push for longer coherence times. 

The article is structure as follows. In Section~\ref{sec:I}, we will provide a short review on the Bose-Hubbard model, its application to the modeling of collectively interacting transmons and the analytical description of the non-Hermitian wavefunctions and energies. This will also allow us to comment on the non-Hermitian relationships between wavefunctions that arise leading to the complications of their dynamics. In Section~\ref{sec:II} we will explore how these non-Hermitian complications can be circumnavigated using anharmonicity as a perturbation and comment the first and second order corrections and their effects on the dark state and its energy. Finally in Section~\ref{sec:III} using these corrections, we will apply the corrections to the master equation and compare the dynamics of the system to established methods which describe non-equilibrium dynamics, namely quenching dynamics and wavefunction relaxation. We highlight at the end that through these methods, a map of transition pathways and correlations between sub- and superradiant states can be built, and show that the transmon pair system naturally relaxes with superradiant-type burst signatures to a stable state. 

%%%%%%%%%%%%%%%%%%%%%%%%%%%%%%%%%%%%%%%%%%%%%%%%%%%%%%%%%%%%%%%%%%%%%%%%%%%%%%%%%%%%%

% -- Section II

%%%%%%%%%%%%%%%%%%%%%%%%%%%%%%%%%%%%%%%%%%%%%%%%%%%%%%%%%%%%%%%%%%%%%%%%%%%%%%%%%%%%%

\section{Non-Hermitian Bose-Hubbard Model }\label{sec:I}
A chain of $L$ photon-emitters can be described by the Bose-Hubbard model written as 
\begin{align}\label{eq:Bose-Hubbard}
    \hat{H}_{\rm BH}/\hbar =  \sum_{j=1}^{\rm L} \omega_j \hat{a}^\dagger_j \hat{a}^{}_j &-\sum_{j=1}^{\rm L} \frac{U_j}{2}\hat{a}^\dagger_j \hat{a}^{}_j\left(\hat{a}^\dagger_j \hat{a}^{}_j-1\right) \nonumber\\ 
    & +\sum_{i,j=1}^{\rm L} J_{ij}\left(\hat{a}^\dagger_i \hat{a}_j + \hat{a}_i \hat{a}^\dagger_j\right),
\end{align}
where $\hat{a}^\dagger_j$ and $\hat{a}_j$ are the harmonic oscillator creation and annihilation operators, $\omega_j$ is the transition frequency on site $j$, $U_j$ is the corresponding on-site interaction strength and $J_{ij}$ is the tunneling rate between sites $i$ and $j$~\cite{Dogra2016,Landig2016,BHReview2008}. In transmons, the values $U_j$ and $J_{ij}$ represent the anharmonicity of each transmon and the capacitive coupling between transmons~\cite{Olli2022,Orell2022}. When describing many-body dynamics, the anharmonicity $U$ acts as a negative on-site interaction which opens interactions between the Fock basis of each site. This staggers the energy difference in each state of a harmonic oscillator bringing them into an anharmonic regime~\cite{Wang2024,Dong2025,Goss2022,Yunzhao2023,Vaaranta2022}, as seen in Fig.~\ref{fig:Schematic}(b). 

The dynamics of this system can be evolved using a master equation. In its evolution, the Bose-Hubbard model is described as an open quantum system, where interactions between individual sub-systems are facilitated through a coupling of a medium such as an open waveguide as seen in Fig.~\ref{fig:Schematic}(a). The waveguide ensures that photon interactions between sites are facilitated while providing a channel in which excess photons, or photons that do not interact between sites, can be dispelled from the system~\cite{Delanty_2011,Greenberg2022}. As an open quantum system the dynamics can be found using the master equation and for a density operator $\hat \rho$ where the evolution of the system is written as
\begin{equation}\label{eq:Master_equation}
    \frac{d\hat \rho }{dt} = -\frac{i}{\hbar}[\hat{H},\hat \rho] + \sum_j \left (  \hat{C}_j\hat \rho \hat{C}_j^\dagger -  \frac{1}{2} \{ \hat{C}_j^\dagger \hat{C}_j,\hat \rho \}\right),
\end{equation}
where the operator $\hat C_j$ is a collective decay operator. Collective operators describe the dissipative relaxation and interactive dynamics of the whole system. For bosonic systems, the operators themselves are written as a superposition of creation and annihilation operators, the multiplication of which leads to the description of the radiative decay and the transfer of excitations between sub-systems dependent on their relative separation,
\begin{equation}\label{eq:Coll_op}
    \hat{C}= \sqrt\frac{\gamma}{L}\sum_{j = 1}^{\rm L} \exp\left(\frac{2\pi i}{\lambda_0}x_j\right)\hat{a}_j.
\end{equation}
Here the exponential term dictates the coherent or incoherent interactions between sites by considering their separation between the $j$th and $k$th transmon, the difference being $|x_j - x_k|$, and $\gamma$ describes the dissipation rate of the system. To ensure all sites interact coherently the distance between each site are set to integer numbers of the photon wavelength $|x_j-x_k| = n\lambda_0$. When applied to the anticommutator term of the master equation, $\{ \hat{C}^\dagger \hat{C},\hat \rho \}$, on-site terms also describe the radiative dissipation while off-diagonal terms track the interactions between sites~\cite{Pichler2015,Shammah2018}.  

In this work, to  ensure that the only form of coupling is of a dissipative nature, the distance between sites is assumed to be large enough such that the effects of tunneling are negligible $J=0$. From these approximations for a pair of transmons, the collective operators simply reduces to the form
\begin{equation}
    \hat{C} = \sqrt\frac{\gamma}{2}(\hat{a}_1 + \hat{a}_2).
\end{equation}
After applying these operators in the master equation, Eq.~\eqref{eq:Master_equation}, the dynamics arising from the anticommutator can be incorporated into the Bose-Hubbard Hamiltonian producing an effective non-Hermitian Bose-Hubbard Hamiltonian written as
\begin{equation}\label{eq:eff_Ham}
    \hat{H}_{\text{eff}}/\hbar = \hat{H}_{\rm BH}/\hbar -\frac{i\gamma}{2}\sum_{i,j}^2 \hat{a}^\dagger_i \hat{a}^{}_j.
\end{equation}
Solutions to the effective Hamiltonian can be found through exact diagonalization, however the complex dissipation term raises concerns for the dynamics and applications to the master equation. The states of the effective Hamiltonian produce complex eigenvalues $\lambda_n = E_n - i\hbar \Gamma_n$ where $E_n$ is the states energy and $\Gamma_n$ is the initial collective decay~\cite{Xiangyu2023}. The effective Hamiltonian is non-Hermitian ~\cite{Hatano1996, Martinez2025}, meaning $\hat H_{\rm eff}\neq \hat H_{\rm eff}^\dagger$ and the states of the Hamiltonian and the Hamiltonian conjugate have to be calculated separately, implying the use of right and left eigenvectors. By themselves, the Hamiltonian is applied to the eigenvectors through
\begin{align}
    \hat{H}_{\rm eff}\ket{\psi^{\rm R}_i} &= \lambda_i\ket{\psi^{\rm R}_i}, & \bra{\psi^{\rm R}_i}\hat{H}_{\rm eff}^\dagger &= \bra{\psi^{\rm R}_i}\lambda_i^*, 
\end{align}
for the right eigenvectors, and similarly
\begin{align}
\hat{H}^\dagger_{\rm eff}\ket{\psi^{\rm L}_i} & =\lambda_i^*\ket{\psi^{\rm L}_i}, & \bra{\psi^{\rm L}_i}\hat{H}_{\rm eff} & = \bra{\psi^{\rm L}_i}\lambda_i,
\end{align} 
for the left eigenvectors. For the calculation of the states in a pair of photon emitters, we find that the left and right eigenvectors form a conjugate relationship with one another simplifying the calculations. When applied to the master equation in Eq.~\eqref{eq:Master_equation} the complex energy allows one to calculate the dissipation rate~\cite{Ordonez2004} to reveal an exponential decay profile. The latter part of the master equation then describes the evolution of a state as it transitions to lower energies through quantum jump events from the $\hat C\hat\rho \hat C^\dagger$ term~\cite{Daley2014,Twanley2019}.

\begin{figure}[t]
    \centering
    \includegraphics[width =\linewidth]{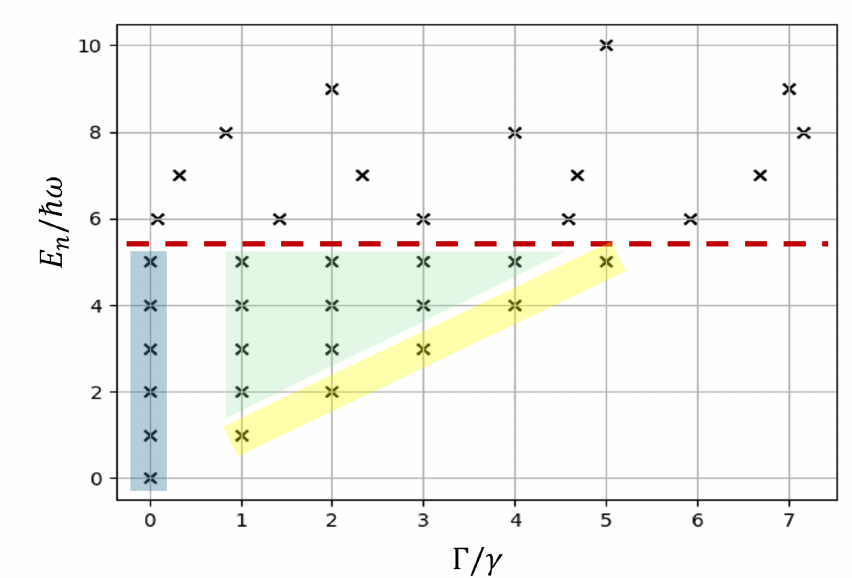}
    \caption{Eigenspectrum of the effective non-Hermitian Hamiltonian of a pair of dissipatively coupled harmonic oscillators $L=2$. Here the local basis is truncated up to maximum $6$ harmonic oscillator states, translated as a maximum excitation of $N_i = 5$ per site. Two regimes can be separated above and below the red dashed line classified as the half-filling line. Below this line, the system is in a harmonic oscillator regime where the eigenspectrum is evenly separated, allowing us to classify the bright (yellow), dark (blue) and faint (green) states. Above this the eigenspectrum globes into a shape reminiscent of a corresponding qubit model.}
    \label{fig:EnergyNoDis}
\end{figure}

\subsection{Harmonic regime with $U=0$}
For a pair of emitters $L=2$ with no on-site interaction $U=0$ the system reduces to a pair of dissipatively coupled harmonic oscillators. Diagonalizing the Hamiltonian we find that the left and right eigenvectors are the same with the only difference being the eigenvalues, which are conjugates of one another. Fig.~\ref{fig:EnergyNoDis} shows the complex eigenspectrum for the right eigenvectors, with the energy on the y-axis and the collective decay on the x-axis. Within each excitation manifold a dark state exists with a dissipation rate of $\Gamma = 0$ and a bright state $\Gamma_{\rm max} = NL\gamma/2$. 

Our simulations consider a local basis of $6$ Fock states revealing a lobed structure to the eigenspectrum. From $N\leq5$, below the red dashed line in Fig.~\ref{fig:EnergyNoDis} the spectrum is evenly spaced on both axis where the total excitation number is below the maximum excitation occupation. A similar lobbing can be seen in qubit arrays, where in collective interacting qubits a dark states exists when up to half of the qubits are excited and at least half are in the ground state. However the collective dynamics of qubit arrays are vastly different to that of the transmon arrays due to their freezing out of higher level excitation~\cite{Zanner2022, Orell2022}.

From the dissipation rate we can determine and classify the states as dark, bright and faint states highlighted in blue, yellow and green respectively in Fig.~\ref{fig:EnergyNoDis}. The wavefunctions form linear combinations of the associated Fock basis of each excitation manifold forming an orthogonal basis. In coupled harmonic oscillator pairs, the wavefunctions also form collective ladder operations through the collapse operators in Eq.~\eqref{eq:Coll_op}~\cite{Celso2025,Orell2022}. These operators differ by a phase of $\pi$ due to the constructive and destructive interference of their interactions and are labeled respectively as dark and bright state operations, written as
\begin{align}
    \hat{d} &= \frac{1}{\sqrt{2}}(\hat{a}_1 - \hat{a}_2), &   \hat{b} &= \frac{1}{\sqrt{2}}(\hat{a}_1 + \hat{a}_2).
\end{align}
When applying these operators one can find symmetric and antisymmetric superpositions of the Fock basis depending on the operator used. Bright state operators preserve the parity of the state, where as dark state operators flip the collective parity. This is important since the parity determines the ability to overlap a state with a perturbing factor such as the anharmonicity~\cite{Greenberg2022,Ordonez2004}.

\subsection{Anharmonic regime}

\begin{figure*}[t]
    \centering
    \includegraphics[width=2\columnwidth]{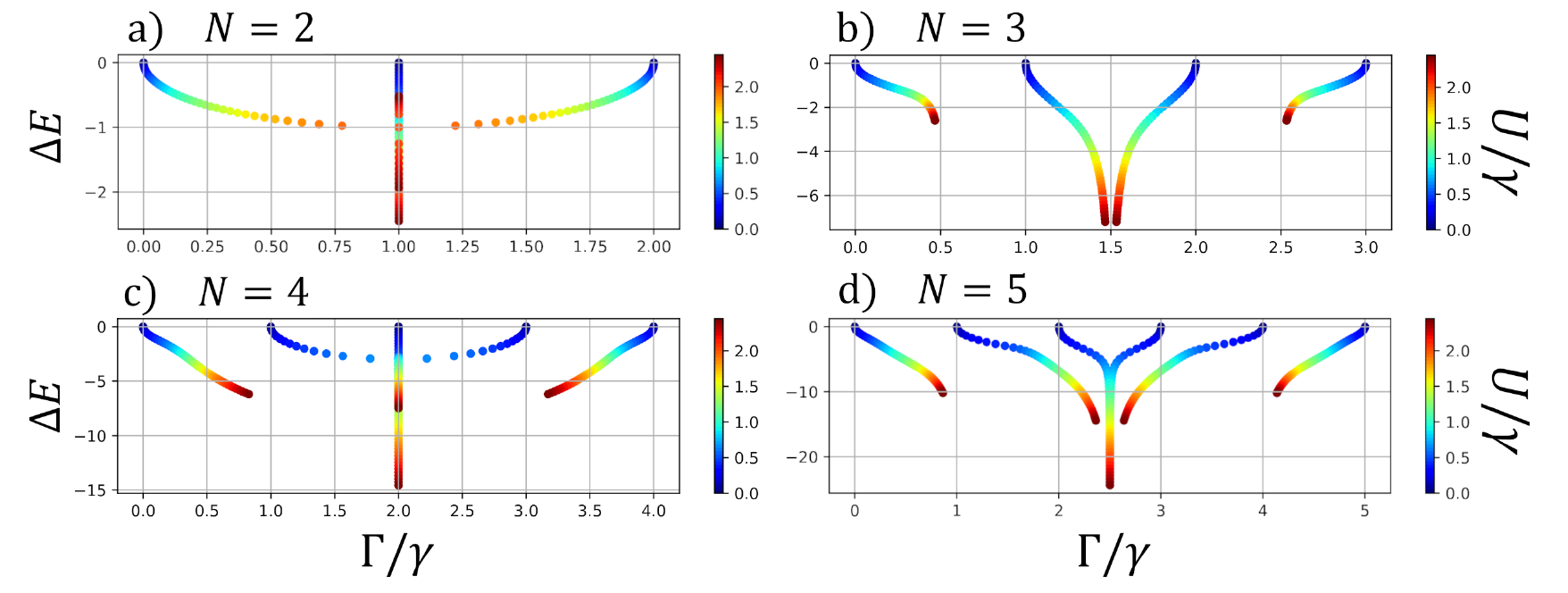}
    \caption{The energy spectrum of the effective non-Hermitian Hamiltonian $\hat H_{\rm eff}$ for $2$ anharmonic oscillators dissipatively coupled as a function of the anharmonicity $U/\gamma \in (0,2.5)$. Panels (a)-(d) are separated in terms of their excitation manifolds from $N=2,3,4,5$. Starting from the harmonic oscillator spectrum (c.f. Fig.~\ref{fig:EnergyNoDis}) an anharmonicity increases the energies coalesce with the behavior dependent on the even or odd number of excitations in the system. }
    \label{fig:Eng_Anharm}
\end{figure*}

Fig.~\ref{fig:Eng_Anharm} presents the numerical evolution of the eigenspectra calculated by exact diagonalization of the right basis as anharmonicity is swept from $U = 0 \xrightarrow{}2.5\gamma$ for the excitation manifolds of $N = 2,3,4,5$ for $L=2$. Immediately, two effects are noticed when anharmonicity is introduced. To start, all states develop some form of dissipation, and the state previously classified as a dark state will become dissipative when anharmonicity is introduced for all excitation manifolds past $N=1$. It can also be seen that for all excitation manifolds, a symmetric coalescence of energies is present, indicating that the effects of dissipation are induced by wavefunction couplings. What differs between manifolds is an even-odd effect, which dictates the behavior of this coalescence. In even excitation numbers, e.g. $N =2,4$, in Figs.~\ref{fig:Eng_Anharm}(a) and~\ref{fig:Eng_Anharm}(c), the coalescence leads to sharp exceptional points around a central faint wavefunction. These artifacts are typical in non-Hermitian eigenspectra and affect the dynamics greatly. In odd excitation numbers, e.g. $N =3,5$, in Figs.~\ref{fig:Eng_Anharm}(b) and Figs.~\ref{fig:Eng_Anharm}(d), coalescence is gradual but still symmetric, with no central wavefunction or definite exceptional point. 

\begin{figure}[tb]
    \centering
    \includegraphics[width =\linewidth]{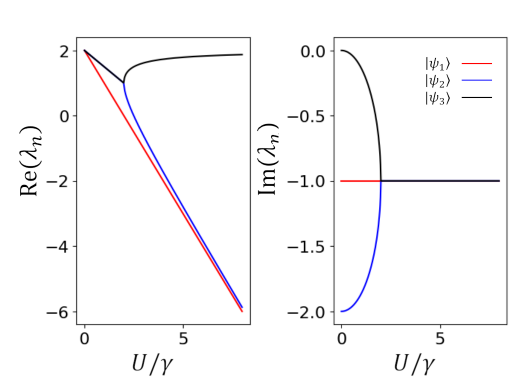}
    \caption{Decomposition of the real and imaginary portions of the energies in Eqs.~(11-13) for the right eigenstates of the effective Hamiltonian for a pair of dissipatively coupled anharmonic oscillators in the $N=2$ excitation manifold as anharmonicity increases. In this figure an exceptional point can be seen at the critical value of $U=2\gamma$ where the energies of the bright and dark states coalesce.}
    \label{fig:N2Energy}
\end{figure}

The anharmonicity acts as an interaction term in the effective Hamiltonian in the perspective of a many-body physics. Importantly, the anharmonicity does not commute with the collective decay term
\begin{equation}
    \left[ \frac{U}{2} \sum_{j=1}^{\rm L} \hat{a}^\dagger_j \hat{a}^{}_j\left(\hat{a}^\dagger_j \hat{a}^{}_j-1\right),\frac{i\gamma}{2}\sum_{i,j=1}^{\rm L}\hat{a}^\dagger_i\hat{a}_j\right] \neq 0.
\end{equation}
This implies that the eigenstates of the effective Hamiltonian are not eigenstates calculated from the collective decay term or of the uncoupled system. Despite this, the Hamiltonian can still be diagonalized with the inclusion of anharmonicity to reveal how non-Hermitian effects change the states and energies. Thus, we can compare the numerical results to the analytical form of the wavefunctions and energies in the $N=2$ excitation manifold. In the Fock basis, the matrix Hamiltonian can be isolated and written as
\begin{equation}
    \hat{H}_{\rm eff}^{N_2}/\hbar = 2\omega \hat I_3 + \left(
\begin{array}{ccc}
 -i \gamma -U & -\frac{i \gamma }{\sqrt{2}} & 0 \\
 -\frac{i \gamma }{\sqrt{2}} & -i \gamma  & -\frac{i \gamma }{\sqrt{2}} \\
 0 & -\frac{i \gamma }{\sqrt{2}} & -i \gamma -U\\
\end{array}
\right),
\end{equation}
where the first term is the harmonic oscillator energy multiplied by an identity matrix $\hat I_3$ and the second term includes all other terms related to anharmonicity and dissipation of the system. The right eigenvalues of the system, plotted separately for the real and imaginary parts in Fig.~\ref{fig:N2Energy}, are written as
\begin{align}
    \lambda_1^{\rm R}/\hbar  =& 2\omega -U-i\gamma,  \\
    \lambda_2^{\rm R}/\hbar  =& 2\omega-\frac{U}{2}-i\gamma -\frac{1}{2}\sqrt{U^2 -4\gamma^2},\\
    \lambda_3^{\rm R}/\hbar  =& 2\omega -\frac{U}{2}-i\gamma +\frac{1}{2}\sqrt{U^2 -4\gamma^2},
\end{align}
with corresponding right wavefunctions
\begin{align}
    \ket{\psi_1^{\rm R}} =& \frac{1}{\sqrt{2}}\left(\ket{20} - \ket{02}\right), \label{eq:analytical_WF1} \\
    \ket{\psi_2^{\rm R}}  =&  A_1\left(\ket{20} +\frac{i \left(U-\sqrt{U^2-4 \gamma ^2}\right)}{\sqrt{2} \gamma }\ket{11} + \ket{02}\right),\\
    \ket{\psi_3^{\rm R}}  =& A_2\left(\ket{20} + \frac{i \left( U+\sqrt{U^2-4 \gamma ^2}\right)}{\sqrt{2} \gamma }\ket{11} + \ket{02}\right), \label{eq:analytical_WF3} 
\end{align}
where $A_i$ is a normalization constant. Immediately it can be seen that at $U=2\gamma$ the wavefunctions $\ket{\psi^{\rm R}_3}$ and $\ket{\psi^{\rm R}_2}$ are indistinguishable and equal, with the corresponding energies also being equal, making $U=2\gamma$ a critical point in the second excitation manifold. For the left eigenvectors, we also find that the wavefunctions and energies are simply conjugates of the right where $\ket{\psi^{\rm L}_i} = \ket{\psi^{\rm R}_i}^*$ and $\lambda_i^{\rm L}=(\lambda_i^{\rm R})^*$. Overlapping left and right eigenvectors for each wavefunction, it can be shown that the overlap between conjugate wavefunctions goes to zero around the critical point due to the complex square root term in the $\ket{11}$ Fock state. This indicates not only a biorthogonal relationship between the two bases, but a point in which the wavefunctions self-orthogonalize~\cite{Fritzsche2026}. This can skew the reliability of the dynamics around and leading to the exceptional point, requiring corrective methods to orthogonalize the system before the initial state is chosen. As exact analytic diagonalization is limited to small system, next we explore alternative, perturbative methods to understand the weakly anharmonic regime.

%%%%%%%%%%%%%%%%%%%%%%%%%%%%%%%%%%%%%%%%%%%%%%%%%%%%%%%%%%%%%%%%%%%%%%%%%%%

%-- SECTION III

%%%%%%%%%%%%%%%%%%%%%%%%%%%%%%%%%%%%%%%%%%%%%%%%%%%%%%%%%%%%%%%%%%%%%%%%%%%

\section{Perturbation Results}\label{sec:II}

In this section we will present a perturbative treatment of the non-Hermitian system described by the effective Bose-Hubbard model. Perturbation methods for non-Hermitian systems have been well studied and new results can extrapolate the power series structure of this method to extend to density operators ~\cite{Sternheim1972,Dolf2023}. Here we focus on the impact of anharmonicity to the originally dark states. In other words, the methods focus on open quantum systems that possess unperturbed steady state solutions, $\frac{d\hat \rho}{dt} = 0$. To construct a perturb density matrix, we start by expanding the density matrix into a power series~\cite{Dolf2023}
\begin{equation}\label{eq:dens_power}
    \hat \rho = \hat \rho_0 +\epsilon\hat\rho_1+\epsilon^2\hat\rho_2  ... 
\end{equation}
where $\hat\rho_0$ is the unperturbed steady state solution and the following operators serve as corrections to this state. Since a density operator is the outer product of a state vector, we can show how each state vector correction contributes to the correction of the density operator using the state vector power series. Starting with,
\begin{equation}
    \ket{\psi} = \ket{\phi_0}+\epsilon\ket{\phi_1}+\epsilon^2\ket{\phi_2} ...
\end{equation}
we construct the full density operator where, 
\begin{align}\label{eq:full_expand}
     \hat \rho=\ket{\psi}\bra{\psi} = &\ket{\phi_0}\bra{\phi_0} + \epsilon\left(\ket{\phi_0}\bra{\phi_1} +  \ket{\phi_1}\bra{\phi_0}\right)  \nonumber \\
     &+ \epsilon^2(\ket{\phi_0}\bra{\phi_2}+\ket{\phi_1}\bra{\phi_1}+\ket{\phi_2}\bra{\phi_0})\nonumber\\
     =& \hat \rho_0 + \epsilon \hat \rho_1 + \epsilon^2 \hat \rho_2,
\end{align}
where the state vector outer products line up with the first and second order corrections to the density operator.

\subsection{First order correction}

Anharmonicity is applied as a perturbation by separating the effective Hamiltonian into its two components. The harmonic and anharmonic portions of the effective Hamiltonian are written as
\begin{align}
    \hat H_0/\hbar  &= \sum_{j=1}^{\rm L}  \omega_j \hat{a}^\dagger_j \hat{a}^{}_j  - \frac{i\gamma}{2}\sum_{i,j}^L \hat{a}^\dagger_i \hat{a}_j.\\
     \hat H_1/\hbar &= -\sum_j^{\rm L}\frac{U}{2}\hat{a}_j^\dagger \hat{a}^{}_j(\hat{a}_j^\dagger \hat{a}^{}_j-1),
\end{align}
where the on-site interaction term acts as a perturbation~$\hat H_1$ and together the full Hamiltonian is written as $\hat H= \hat H_0+\epsilon \hat H_1$. Starting from our pair of of dissipatively coupled harmonic oscillators, the dark state in the $N=2$ excitation manifold is written as 
\begin{equation}
    \ket{\psi_{\rm DS}} = \frac{1}{2} \left(\ket{20} + \ket{02}\right) - \frac{1}{\sqrt{2}}\ket{11}.
\end{equation}
The basis for the whole system is orthogonal with the bright and fain state. The eigenvectors are found using exact diagonalization methods of the effective Bose-Hubbard Hamiltonian in Eq.~\eqref{eq:eff_Ham} when $U=0$. We also note that the eigenvectors are the same on the left and right eigenvectors, mitigating the need for these consideration for now.

For the first order correction to the energy, the dark state is simply overlapped with the perturbing Hamiltonian
\begin{equation}
    E_1/\hbar = \bra{\psi_{\rm DS}}\hat H_1\ket{\psi_{\rm DS}} = -\frac{U}{2} \ .
\end{equation}
For the first order correction of the wavefunction the overlap of the dark state with all other states in its excitation manifold are calculated and summed over its basis. Since the eigenspectrum can be separated into real and imaginary components the correction can be calculated using non-degenerate perturbation theory using expectation values of the effective Hamiltonian in the same excitation manifold,
\begin{equation}
    \ket{\phi_1} = -\sum_{n\neq0}  \frac{ \bra{\psi_n}\hat H_1\ket{\psi_{\rm DS}}}{\left(\langle \hat H_0\rangle_n-E_{\rm DS}\right)} \ket{\psi_n},
\end{equation} 
where $\langle\hat H_0\rangle_n$ is energy expectation value of the perturbing wavefunction when $U=0$, $E_{\rm DS} = 2\hbar\omega$ is the energy of the dark state and $\ket{\psi_n}$ is the basis of the harmonic oscillator bright, dark and faint states. Due to the collective symmetries of the wavefunctions the dark state only overlaps with the bright state in the $N=2$ excitation manifold, and conversely for the antisymmetric nature of the faint state $\ket{\psi_{\rm F}} = \frac{1}{\sqrt 2}(\ket{20}-\ket{02})$, the overlap with the dark state goes to zero. This then leads to the calculation of the first order correction being, 
\begin{equation}\label{eq:FO_correction}
    \ket{\phi_1} = \frac{iU}{4\gamma}\ket{\psi_{\rm BS}},
\end{equation}
where the bright state $\ket{\psi_{\rm BS}}$ is written as
\begin{equation}
    \ket{\psi_{\rm BS}} = \frac{1}{2}(\ket{20}+\ket{02})+\frac{1}{\sqrt 2}\ket{11} \ .
\end{equation}
Once the full wavefunction is constructed as a product state the first order correction to the density operator can be found
\begin{equation}\label{eq:correlation}
   \hat \rho_1 = \frac{iU}{4\gamma}\left(\ket{\psi_{\rm DS}}\bra{\psi_{\rm BS}} - \ket{\psi_{\rm BS}}\bra{\psi_{\rm DS}}\right) \ .
\end{equation}
This indicates that in the presence of anharmonicity, a complex valued and symmetric correlation forms between bright and dark states at the $N=2$ manifold. The formation of this correlation opens a transition path in which a dark states population can transition into a dissipative state allowing the population of the system to dissipate into the ground state. 

This can be shown for all dark states where the nearest neighbor similarly symmetric state, more specifically the state that exhibits an initial dissipation rate of $2\gamma$, is the only state that couples to the dark state. For the $N=3$ manifold, this means that the antisymmetric dark state will couple to the nearest anti-symmetric state. Moving into higher manifolds, these correlations result in a cascading expulsion of photons in their relaxation, where states couple via off diagonal correlations. Interestingly dark states initialized with an odd number of excitations relax into the dark state of the $N=1$ manifold, due to their antisymmetric parity and the first excitation dark state being the remaining dark one in the presence of anharmonicity. Likewise, even excitations then relax into the ground state, which will examined in detail in next section.

\subsection{Second order correction}
\begin{figure}
    \centering
    \includegraphics[width=\columnwidth]{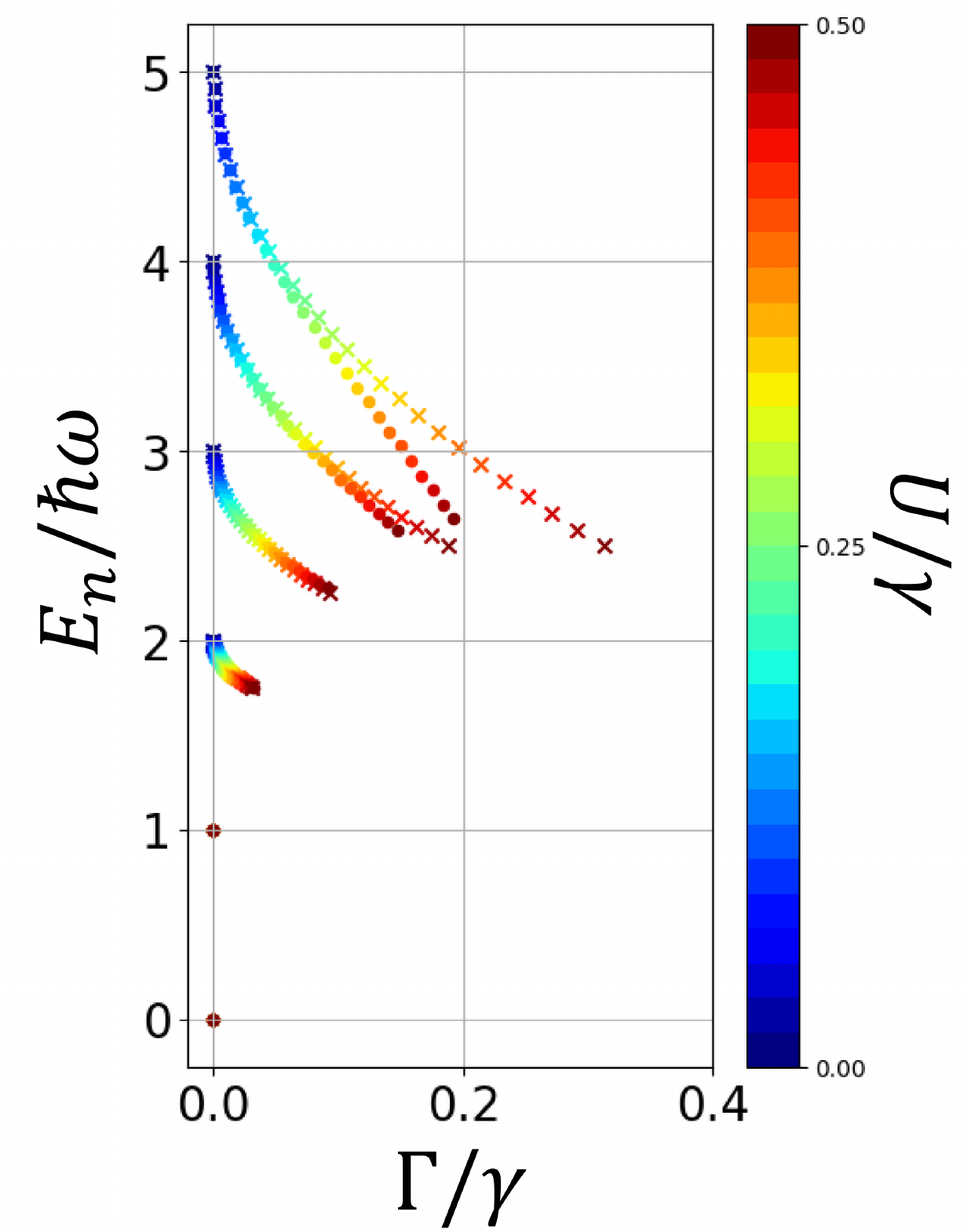}
    \caption{
    Analytical perturbative correction calculated from Eq.~\eqref{eq:all_correction} compared to the energy spectrum for the dark states of energy excitation manifolds $N = 0\xrightarrow{}6$. Analytical results (crosses) are compared to a small section of the numerical results (c.f. Fig.~\ref{fig:Eng_Anharm}) (dots) for small perturbations where $U/\gamma \in (0,0.5)$}
    \label{fig:pert}
\end{figure}

With the calculation of the first order correction to the wavefunction, the second order energy can be calculated using the overlap of the unperturbed wavefunction with the first order wavefunction correction, $\bra{\psi_{\rm DS}}\hat H_1\ket{\phi_1}$. The total correction of the energy for the second excitation manifold can the be written as
\begin{equation}
    \lambda_c(N=2) = -\frac{\hbar U}{2} - \frac{i\hbar U^2}{8\gamma}.
\end{equation}
The second order correction indicates a shift in the imaginary portion of the energy inducing a dissipation that is directly linked to the induced correlation between the dark state and its neighboring dissipative state. This can be generalized to all dark state in each excitation manifold, where the correction is written in terms of the excitation number as $N$
\begin{equation}\label{eq:all_correction}
    \lambda_c(N) = \frac{N(N-1)\hbar U}{4} -\frac{iN(N-1)\hbar U^2}{16\gamma}.
\end{equation}
The analytical corrections for the energies are plotted in Fig.~\ref{fig:pert} compared to the numerical results up to the fifth excitation manifold and up to an anharmonicity of $U/\gamma=0.5$. As the number of excitations increase, the analytical results breakdown faster, with low excitation numbers having close agreement to the numerical results. Importantly, all energy corrections point to an induced dissipation in the system in the presence of anharmonicity.

Following the work of Huybrechts and Roscilde in Ref.~\onlinecite{Dolf2023}, the second order wavefunction correction can be calculated by extending the perturbations to include the jump operators. In their derivation, modified to our analysis of dark states, the general second order correction takes the form of
\begin{align}\label{eq:second_order_general}
    \ket{\phi_2} = \sum_{n,m\neq0}&\frac{\bra{\psi_n}\hat H_1\ket{\psi_m}\bra{\psi_m}\hat H_1\ket{\psi_{\rm DS}}}{(\langle \hat H_0^{(NH)}\rangle_n-E_{\rm DS})(\langle \hat H_0^{(NH)}\rangle_m-E_{\rm DS})}\ket{\psi_n} \nonumber \\ 
    &-\sum_{n\neq0}\frac{\bra{\psi_{\rm DS}}\hat H_1\ket{\psi_{\rm DS}}\bra{\psi_n}\hat H_1\ket{\psi_{\rm DS}}}{(\langle \hat H_0^{(NH)}\rangle_n-E_{\rm DS})^2}\ket{\psi_n} \nonumber \\
    &-\frac{1}{2}\sum_{n\neq0}\frac{|\bra{\psi_n}\hat H_1\ket{\psi_{\rm DS}}|^2}{|\langle \hat H_0^{(NH)}\rangle_n-E_{\rm DS}|^2}\ket{\psi_{\rm DS}}  \\
    &-\sum_j \gamma_j \sum_{n\neq0}\frac{\bra{\psi_n}\hat{C}_j\ket{\phi_1}\bra{\phi_1}\hat{C}_j^\dagger\ket{\psi_{\rm DS}}}{\langle \hat H_0^{(NH)}\rangle_n-E_{\rm DS}}\ket{\psi_n}.  \nonumber
\end{align}
When initialized to a specific manifold the final term can be neglected since applying collective operator pertains to neighboring excitation manifolds. We also see that in the $N=2$ example, the first two terms also cancel with one another since we only consider a manifold with three states and $\bra{\phi_{\rm BS}}\hat H_1\ket{\phi_{\rm BS}} =  \bra{\phi_{\rm DS}}\hat H_1\ket{\psi_{\rm DS}}$. This leaves us with the final term where the second order wavefunction correction can be written as
\begin{equation}
    \ket{\phi_2} = -\frac{1}{2}\left(\frac{U}{4\gamma}\right)^2\ket{\psi_{\rm DS}}.
\end{equation}
The second order correction to the density operator can then be written as
\begin{equation}\label{eq:second_correction}
    \hat \rho_2 = \left(\frac{U}{4\gamma}\right)^2(-\ket{\psi_{\rm DS}}\bra{\psi_{\rm DS}} + \ket{\psi_{\rm BS}}\bra{\psi_{\rm BS}}) \ .
\end{equation}
We now see that the second order correction indicates a transition from a dark state solution to a state that has both dark and bright state components. This is evident from the induced dissipation of the total correction to the energy, since an initial dissipation rate is induced as anharmonicity is introduced in Eq.~\eqref{eq:all_correction}, seen at $E_n/\hbar\omega = 2$ in Fig.~\ref{fig:pert}. With part of the population now shifting to the bright state in its initialization, the state now has a portion which can readily expel photons from system. Since we want to include an analysis of the dynamics we have focused mainly on the dark state; however, if the same perturbative treatment was done on the bright state, we would find the same induced correlation and state mixing as with the dark state.

\section{Dynamics Stability}\label{sec:III}

In this section we will compare the dynamics of the exact numerical simulation and the perturbation method starting from the dark state as the initial state, using the master equation in Eq.~\eqref{eq:Master_equation}. Numerically evolving an eigenstate from a non-Hermitian Hamiltonian can be complicated when biorthogonal properties of left and right eigenvectors need to be taken into consideration. In our system, partial overlaps are also present between wavefunctions, in particular between the $N=2$ bright and dark states. In order to initialize the dark state we need to use a method that both renormalizes and orthogonalizes the bases simultaneously. To do this, we use the Gram-Schmidt orthogonalization method. For our left and right eigenvectors, this algorithm must be implemented simultaneously as the corrections feed into each other in the following way
\begin{align}
    \ket{\phi^{\rm R}_k} &= \ket{\psi^{\rm R}_k} - \sum_{j=1}^{k-1}\frac{\bra{\phi^{\rm L}_j}\ket{\psi_k^{\rm R}}}{\bra{\phi^{\rm L}_j}\ket{\phi_j^{\rm R}}} \ket{\phi^{\rm R}_j}, \\
    \ket{\phi^{\rm L}_k} &= \ket{\psi^{\rm L}_k} - \sum_{j=1}^{k-1}\frac{\bra{\phi^{\rm R}_j}\ket{\psi_k^{\rm L}}}{\bra{\phi^{\rm R}_j}\ket{\phi_j^{\rm L}}} \ket{\phi^{\rm L}_j}.
\end{align}
This makes the order in which the corrections are applied important since the correction of each state depends on the previously corrected states. Therefore, for the $N=2$ manifold, we start with the asymmetric faint state $\ket{\psi^{R(L)}_1}$ since there is no change in the state vector with the introduction of anharmonicity. We end on the state $\ket{\psi_3^{R(L)}}$ since at $U=0$ the state reduces to the harmonic dark state. Through the calculation, we see that the only correction in this algorithm is that of $\ket{\psi^{R(L)}_2}$ allowing us to compare the numerical evolution to the corrected wavefunction found from the perturbation calculations in Sec.~\ref{sec:II}.

From the corrections calculated from the perturbation methods presented, the corrected density matrix can be directly applied to the master equation. The advantage of this approximation is through the utilization of the orthogonal harmonic oscillator basis. For a density operator expanded in a power series using $\hat \rho = \ket{\psi}\bra{\psi}$, a power series for the master equation can be constructed using $\hat H=\hat H_0+\epsilon \hat H_1$. From here the master equation splits into a series of effective Hamiltonians and Liouvllian functions that describe the dynamics of the state once fully expanded 
\begin{align}
    \dot{\hat \rho} =& -\frac{i}{\hbar}[\hat H_0 + \epsilon \hat H_1,(\hat \rho_0 + \epsilon\hat \rho_1+\epsilon^2\hat \rho_2)]  - \frac{\gamma}{2}\sum \hat C_i\hat \rho \hat C_i^\dagger\nonumber \\
    =& -\frac{i}{\hbar}[\hat H_0,\hat \rho_0]-\frac{i}{\hbar}\epsilon[\hat H_1,\hat \rho_0]-\frac{i}{\hbar}\epsilon[\hat H_0,\hat \rho_1] ... \nonumber \\
    &...  -\frac{i}{\hbar}\epsilon^2[\hat H_0,\hat \rho_2] -\frac{i}{\hbar}\epsilon^2[\hat H_1,\hat \rho_1] - \frac{\gamma}{2}\sum \hat C_i\hat \rho \hat C_i^\dagger
\end{align}
where $\hat \rho_1$ and $\hat \rho_2$ are the first and second order corrections calculated in Eqs.~\eqref{eq:correlation}~and~\eqref{eq:second_correction} and $\hat{C}^{(\dagger)}$ are the associated collapse operators. For our corrected $N=2$ dark state density operator we immediately notice the first term goes to zero with all other values not going to zero. This indicates that the method will remain consistent in that an induced initial dissipation will be seen with a continuous decay in the relaxation. 
 
\begin{figure}
    \centering
    \includegraphics[width=\columnwidth]{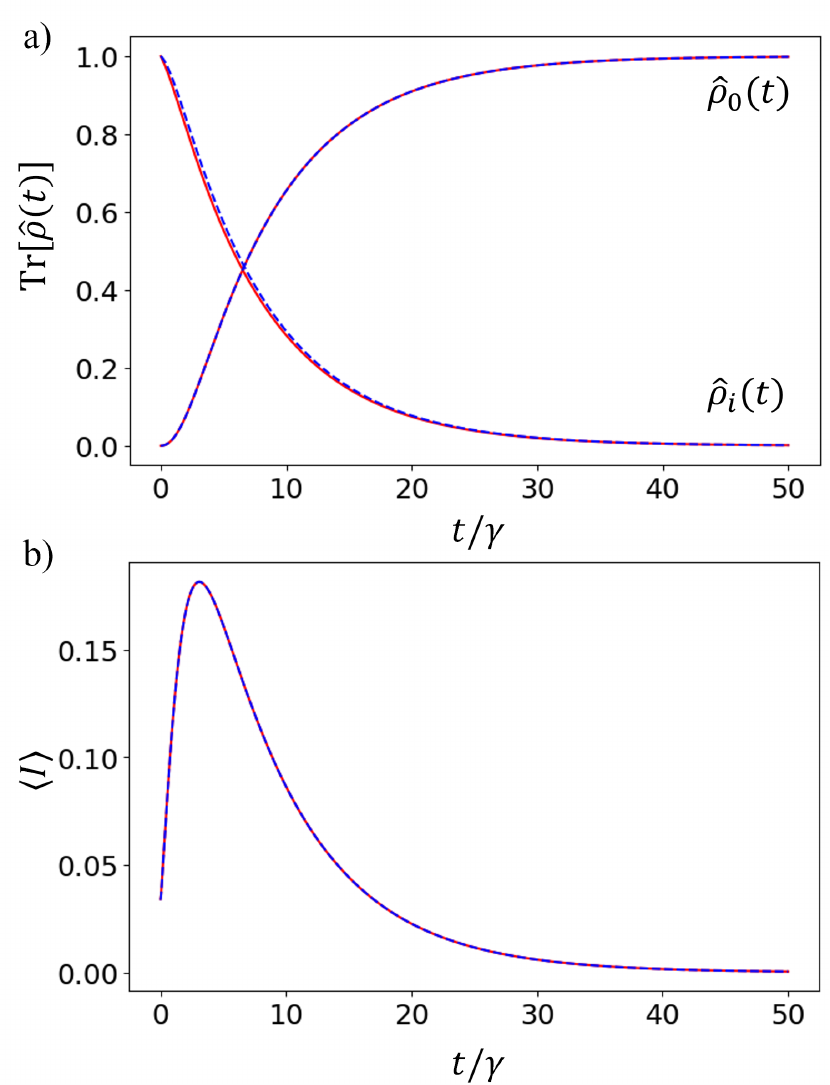}
    \caption{Numerical (red solid line) vs. Perturbation (blue dotted line) dynamics for a dark state initialized with an anharmonicity of $U/\gamma=0.5$. a) Dynamics of the relaxation dynamics of the population of the respective initial states starting from $\hat \rho_i$ relaxing into the ground state $\hat \rho_0 = \ket{00}\bra{00}$ over time. In the numerical evolution, the density matrix is normalized at each time step as is standard for non-Hermitian evolutions. b) The calculated photon readout $\langle I \rangle$. A clear initial burst of photons can be seen in each case at the start of the evolution.}
    \label{fig:Superradiance}
\end{figure}
Fig.~\ref{fig:Superradiance}(a) shows the dynamical relaxation of $N=2$ dark state that has been computed from exact numerical simulations (blue) and through our proposed perturbative methods (red). In Fig.~\ref{fig:Superradiance}(a), the dark state in both cases clearly exhibits decay with an incremental speed up over a short time to a near exponential decay throughout the evolution. Comparing their rates of decay there is a negligible difference between the numerical and perturbative results and both can be seen relaxing into the ground state $\hat\rho_0$ at similar rates. This shows a close agreement between the perturbative corrections with commonly used numerical methods for small perturbations.  Fig.~\ref{fig:Superradiance}(b) then plots the photon read-out of the resulting evolution, more specifically the rate of change of the total excitation number of the system. In a lossy waveguide excitations are expelled from the system. The rate of this expulsion is calculated by summing over the rate of change in each excitation manifold as the system relaxes, explicitly
\begin{equation}
    \langle I\rangle(t) = \frac{d}{dt}\sum_k \langle \hat a_k^\dagger \hat a_k \rangle(t) \, ,
\end{equation}
where $\langle \hat a^\dagger_k \hat a_k\rangle$ is the expectation value of the excitation in the $k$th manifold. The results between the analytical and numerical dynamics closely match one another meaning that measurements between excitation manifolds remain unchanged. 

The advantage of the perturbative approach allows one to conceive a map of transition events during the relaxation of from the excited state. From our corrections of the wavefunctions at $t=0$ in Fig.~\ref{fig:Superradiance}(b), an initial intensity is seen due to the mixing of populations between the dark and bright density operators. Since the bright state is by nature maximally dissipative in the $N=2$ manifold, the population will immediately start to decay over time into the $N=1$ excitation when initialized according to the second order correction to the density operator. Subsequently, we see the population replenished due to the correlation between the dark and bright state in the first order density operator correction. 

Surprisingly in both cases, since the rate in which population will travel to the bright state exceeds the rate in which the state decays, the measurement of the photon intensity increases into a burst resembling to that of superradiance. In the perturbative case this can be mapped for all dark states in the higher excitation manifolds, where the transition events form a cascading decay of photons into states of varying dissipation rates. We note once again that for the dynamics of all dark states, the final state these states will decay into is dependent of the initialization between even and odd excitation numbers. Dark states of even numbers of excitations will decay into the ground state, but odd excitations will decay and remain trapped in the first excitation dark state. This is attributed to transitions being parity preserving actions on top of the fact that the first excitation manifold states do not change in the presence of anharmonicity in the Bose Hubbard model in Eq.~\eqref{eq:Bose-Hubbard}.

\section{Discussions}

While closely following the results of the numerical simulations for a weak anharmonicity we note that the analytical corrections can be shown to deviate from the predicted numerical values for stronger anharmonicties near and beyond the exceptional point. This is to no surprise since perturbation methods typically rely on weak perturbations. The methods presented are still beneficial since the time evolution of the system is numerically stable, trace preserving and allows for a complete tracking of the population of the system when contrasted to the numerical results. By having a full map of the transitions and state populations, population dynamics can be predicted and acted upon, especially in quantum control schemes for superconducting circuits~\cite{Kumar2016,Serra2026,Singhal2025}. If these perturbative results provide enough information to achieve continuous population control, then the next step would be to investigate other dephasing and dissipating factors in the system, especially in qubit models where there is more interest in information retention and control.  

In terms of the effect of anharmonicity specifically, another area to continue this analysis is in scalability by increasing the number of subsystems considered and find ways to simplify the calculations to show how the effects of perturbations scale with the system size. We have purposefully chosen to focus on a small system such as a pair of anharmonic oscillators to highlight the technical challenges faced in the context of non-Hermitian systems. The system also conveniently posses dark states in a non-degenerate orthogonal basis. By increasing the number of pairs, the global interactions of the system also have to be taken into consideration, and will no doubt require methods of degenerate and non-degenerate perturbative methods. However by reformulating the collective interactions of a global system as a collection of local pair operators we believe that the full dynamics of a system can be tracked through these methods, as can be seen through the modeling of spin dimers~\cite{Pichler2015,Pichler2014}.

\section{Conclusions}
The method presented focuses on analyzing anharmonicity as a perturbation and extrapolating the effects of anharmonicity through the orthogonal basis of two dissipatively coupled harmonic oscillators. From our results, we showed that anharmonicity induces dissipation in the dark state primarily through an induced coupling with a dissipative state with similar collective parity. For all dark states, this allows one to construct a map of the transfer of populations during the relaxation of the dark states into the ground state. This can be verified by constructing density operators for the corrected dark state and applying this operator to the master equation, which is a steady state solution in the harmonic oscillator basis. By testing these methods against already established numerical methods of the dynamics, we confirmed that the evolution of both systems is within close agreement with one another.

In principle, by applying the perturbed density matrix to the master equation, the dynamics of the initialized system can be used to calculate other parameters such as coherence times of higher order dark states. In doing so methods to counteract and suppress the destruction of dark states in dephasing and driving protocols can be carefully examined, pushing for higher state fidelities in these processes. This would also be an interesting venture for development of gate protocols in the qudit regime and push the viability of information structures past the qubit level.

\acknowledgements

This work was supported by the Research Council of Finland by Grant Nos. 355824 and 352788 H2Future and the  University of Oulu. We also want to thank  Gonzalo Mart\'in V\'azquez for their initial discussions on the topic and Anna Khylenko for their input in the ongoing work.

\bibliography{biblo.bib}

\end{document}